\documentclass[12pt,a4paper]{iopart}

\usepackage{amsfonts,iopams,cite}

\usepackage{bbm,mathrsfs}
\usepackage{graphicx}
\usepackage{dcolumn}
\usepackage{psfrag}

\newcommand{\id}{\mathbbm{1}}
\newcommand{\im}{\mathbbm{i}}
\newcommand{\ket}[1]{\vert #1 \rangle}

\newcommand{\meanv}[1]{\big\langle #1 \big\rangle}
\newcommand{\var}[1]{\left( \Delta #1\right)^2}
\newcommand{\varsb}[1]{\left[ \Delta #1\right]^2}

\begin{document}

\title{ A continuous-variable formalism for the Faraday atom--light interface}
\date{\today}

\author{Julia Stasi\'nska$^{1,\#}$, Simone Paganelli$^1$, Carles Rod\'o$^1$ and Anna Sanpera$^{2,1}$}

\address{$^1$ Grup de F\'isica Te\`orica: Informaci\'o i Fen\`omens Qu\`antics,\\
 Universitat Aut\`onoma de Barcelona, 08193 Bellaterra (Barcelona), Spain.}




\address{$^2$ ICREA-Instituci\'o Catalana de Recerca i Estudis Avan\c cats.\\
Lluis Companys 23, 08010 Barcelona, Spain}

\eads{ \mailto{$^{\#}$ julsta@ifae.es}}

\begin{abstract}
Quantum interfaces between polarized atomic ensembles and coherent states of
light, applied recently to manipulate bipartite and multipartite entanglement,
are revisited by means of a continuous-variable formalism. The explicit use of
the continuous-variable formalism facilitates significantly the analysis of
entanglement between different modes, reducing it to the study of the
properties of a final covariance matrix which can be found analytically.
Furthermore, it allows to study matter--light interfaces for mixed states,
adapting the formalism to the experimental situations in which the initial
prepared Gaussian states are, unavoidably, affected by a certain amount of
noise. A multipartite scenario, leading to the generation of macroscopic
cluster states is presented and analyzed in detail within this formalism.

\end{abstract}




\section{Introduction}\label{intro}
A strongly polarized macroscopic atomic ensemble can be described disregarding
its individual atomic components and using collective spin variables instead.
In this scenario, the interaction of an atomic polarized ensemble with
off-resonant linearly polarized light resolves, at a classical level, into a
Faraday rotation. The latter refers to the rotation experienced by the
polarization of light propagating inside a magnetic medium (polarized atomic
ensemble). Moreover, this interaction can also lead to an exchange of quantum
fluctuations between matter and light, i.e. a quantum interface. Seminal
results exploiting such interface are the generation of spin squeezed states
\cite{Kuzmich1997_spinsq}, and the entanglement between two macroscopic
spatially-separated atomic ensembles \cite{Julsgaard2001_Nat}. In the first
case, squeezing was produced by the interaction of an atomic sample with a
squeezed state of light, whereas the second task was achieved by propagating a
single laser bram through both samples. The above protocols demand a final
measurement on the light (homodyne detection) which projects the state of the
atomic ensembles into a desired final state. This mapping of fluctuations
between different physical systems provides a powerful tool to design quantum
correlations. The potential applications of such a genuine quantum instrument
have just started. Among them the use as a spectroscopic method for strongly
correlated ultracold atomic systems \cite{Sorensen1998_spectroscopy}.
Furthermore, if the incident light beam is spatially tailored, the interface
provides spatial resolution. This characteristic should permit to directly
observe correlations without the need of individual spin addressing
\cite{Eckert2007_nphys,Roscilde2009_polspectr,Bruun2009_spincor}, or even
provide a direct measure of the order parameter of some exotic quantum states
of matter \cite{deChiara2010}.  Conversely, a quantum interface between a
strongly correlated system and light should allow to map the correlations of
the matter onto the light, leading to the generation of highly non classical
states of light.

Here we address matter--light interfaces in the framework of
continuous-variable (CV) formalism which strongly facilitates the manipulation
and verification of entanglement, both bipartite and multipartite, reducing it
to the study of the properties of a final covariance matrix. Such formalism is
especially well suited for mixed states since they are treated on an equal
footing as pure states. The formalism of continuous variables arises naturally
if the initial states of atomic ensembles and light are Gaussian. Then, if the
interaction between light and matter is bilinear, the Gaussian character of the
subsystems is preserved. Unitary evolutions of such form correspond to
symplectic transformations of the covariance matrix of the composite system
(matter and light). The final projective measurement on the light, homodyne
detection, is also a Gaussian operation and it is easily performed at the level
of the final covariance matrix. Therefore, all the necessary tools are at hand.

The paper is organized as follows. In section 2, we briefly state the problem
under consideration, starting from the description of the matter and light as
continuous-variable systems. Then using the Quantum Non Demolition (QND)
effective Hamiltonian of the matter--light interaction we derive the evolution
equations (for a detailed derivation of such Hamiltonian the reader is referred
to \cite{Julsgaard_PhD}). In section 3, we derive explicitly all the steps of
the matter--light interface in terms of covariance matrices and symplectic
transformations \cite{Madsen2004_Gauss,Sherson2005_Gauss1}. In section 4, we
demonstrate the suitability of the formalism by analyzing in detail two
examples. The first one corresponds to the well known EPR bipartite
entanglement between two macroscopic atomic ensembles, experimentally
demonstrated in \cite{Julsgaard2001_Nat}. We consider a more general scenario
in which the atomic samples are initially in thermal states with different
amount of noise. In such a case entanglement between the atomic samples is
produced only above a given value of the matter--light coupling constant. By
exploiting different geometries \cite{Stasinska2009_meso}, the interaction with
a second light beam permits to design optimally the final entanglement between
the samples and even to delete the entangling action of the first beam. The
second case we treat, much more involved, analyzes the generation of cluster
states using matter--light interfaces and demonstrates the full power of this
formalism in the multipartite scenario. In section 5, we present our
conclusions.

\section{The Faraday Interaction}

The physical system we analyze consists of several atomic ensembles and light
beams, the latter playing the role of information carriers between the atomic
samples. At a time, only a single light beam interacts with the atomic
ensembles. After the interaction, the light beam is measured.

Each atomic ensemble contains a sufficiently large number, $N_{\mathrm{at}}$,
of noninteracting alkali atoms with individual total angular momentum
$\mathbf{\hat{F}}$. The ensemble is described by its collective angular
momentum $\mathbf{\hat{J}}=(\hat{J}_{x},\hat{J}_{y},\hat{J}_{z})$, where
$\hat{J}_{k}=\sum_{i=1}^{N_{\mathrm{at}}}\hat{F}_{k,i}$ ($k=x,y,z$). All atoms
are assumed to be polarized along the $x$ direction, which corresponds to
preparing them in a particular hyperfine state $\ket{F,m_F}$. In such a
situation, fluctuations in the $\hat{J}_x$ component of the collective spin are
very low, allowing for treating this variable as a classical number
$\hat{J}_x\approx \meanv{\hat{J}_x}\equiv \hbar J_x=\hbar N_{\mathrm{at}} F$.
As a consequence, the quantum character of the collective spin is preserved in
the orthogonal spin components, which have a zero mean, but non-zero
fluctuations. By appropriate normalization they are made to fulfil the
canonical commutation relation, $\left[\hat{J}_y/\sqrt{\hbar
J_x},\hat{J}_z/\sqrt{\hbar J_x}\right]=\im \hbar$. To stress the continuous
variable character of the system, we rename the above variables as ``position''
and ``momentum'' :
\begin{eqnarray}\label{canonical_at}
\hat{x}_{A}&=&\frac{\hat{J}_{y}}{\sqrt{\hbar J_{x}}},\nonumber\\
\hat{p}_{A}&=&\frac{\hat{J}_{z}}{\sqrt{\hbar J_{x}}},
\end{eqnarray}
and from now on use only the canonical variables $\hat{x}_{A}, \hat{p}_{A}$ to
refer to the atomic sample, where the subindex $A$ stands for atomic ensemble.
Later on when we deal with few ensembles we will use the notation
$\hat{x}_{A,n}, \hat{p}_{A,n}$ to refer to the $n$th atomic sample.

On the other hand, the light is taken to be out of resonance from any relevant
atomic transition and linearly polarized along, say, the $x$-direction. We use
the Stokes vector description
$\mathbf{\hat{s}}=(\hat{s}_x,\hat{s}_y,\hat{s}_z)$ of light polarization. The
components $\hat{s}_k$ $(k=x,y,z)$ correspond to the differences between the
number of photons (per unit time) with $x$ and $y$ linear polarizations, $\pm
\pi/4$ linear polarizations and the two circular polarizations, i.e.
\begin{equation}\label{stokes}
\left\{
\begin{array}{l}
\displaystyle \hat{s}_x=\frac{\hbar}{2} (\hat{n}_x-\hat{n}_y)=\frac{\hbar}{2} (\hat a_x^\dag\hat a_x-\hat a_y^\dag\hat a_y), \\[2ex]
\displaystyle \hat{s}_y=\frac{\hbar}{2} (\hat{n}_{\nearrow}-\hat{n}_{\searrow})=\frac{\hbar}{2}(\hat a_x^\dag\hat a_y+\hat a_y^\dag\hat a_x),\\[2ex]
\displaystyle
\hat{s}_z=\frac{\hbar}{2}(\hat{n}_{\circlearrowleft}-\hat{n}_{\circlearrowright})=\frac{\hbar}{2 \im}(\hat a_x^\dag\hat a_y-\hat a_y^\dag\hat a_x)
\end{array}\right.
\end{equation}

These allow for the microscopic description of the interaction with atoms,
however, effectively only the following macroscopic observables will be
relevant: $\hat{S}_k=\int_{0}^{T} \hat{s}_k(t)\mathrm{d}t$, where $T$ is the
duration of the light pulse. So defined operators obey standard angular
momentum commutation rules. The assumption of linear polarization along
direction $x$ allows for the approximation
$\hat{S}_x\approx\meanv{\hat{S}_x}\equiv N_{ph} \hbar/2 $. Once more, the
remaining orthogonal components $\hat{S}_y$ and $\hat{S}_z$ are appropriately
rescaled in order to make them to fulfill the canonical commutation rule,
$\left[\hat{S}_y/\sqrt{\hbar S_x},\hat{S}_z/\sqrt{\hbar S_x}\right]=\im\hbar$.
Straightforwardly, a correspondence equivalent to equation (\ref{canonical_at})
arises:
\begin{eqnarray}\label{canonical_light}
\hat{x}_{L}&=&\frac{\hat{S}_{y}}{\sqrt{\hbar S_{x}}},\nonumber\\
\hat{p}_{L}&=&\frac{\hat{S}_{z}}{\sqrt{\hbar S_{x}}},
\end{eqnarray}
which allows to treat the light polarization degrees of freedom on the same
footing as the atomic variables. Notice, that while only one light beam is the
carrier which entangles the atomic ensembles, a secondary light beam will be
normally needed to verify entanglement \cite{Julsgaard_PhD,Stasinska2009_meso}.

In the situation in which a light beam propagates in the $YZ$ plain and passes
through a single ensemble at angle $\alpha$ with respect to direction $z$, the
atom-light interaction can be approximated to the following QND effective
Hamiltonian (see \cite{Julsgaard_PhD} for a detailed derivation)

\begin{equation}\label{hamiltonian}
\hat{H}_{\mathrm{int}}^{\mathrm{eff}}(\alpha)=-\frac{\kappa}{T} \hat{p}_{L}
(\hat{p}_{A} \cos{\alpha}+\hat{x}_{A} \sin{\alpha}).
\end{equation}
The parameter $\kappa$ is the coupling constant with the dimension of the
inverse of an action. Notice that such Hamiltonian leads to a bilinear coupling
between the Stokes operator and the collective atomic spin operators. Evolution
can be calculated through the Heisenberg equation for the atoms and using
Maxwell--Bloch equation for light, neglecting retardation effect. The variables
characterizing the composite system transform according to the following
equations (\cite{Julsgaard_PhD} and references therein):
\numparts
\begin{eqnarray}
\hat{x}_{A}^{\mathrm{out}}&=&\hat{x}_{A}^{\mathrm{in}}-\kappa \hat{p}_L^{\mathrm{in}} \cos \alpha,\label{propagationa}\\
\hat{p}_{A}^{\mathrm{out}}&=&\hat{p}_{A}^{\mathrm{in}}+\kappa \hat{p}_L^{\mathrm{in}} \sin \alpha,\label{propagationb}\\
\hat{x}_L^{\mathrm{out}}&=&\hat{x}_L^{\mathrm{in}}-\kappa (\hat{p}_{A}^{\mathrm{in}} \cos{\alpha}+\hat{x}_{A}^{\mathrm{in}} \sin{\alpha}),\label{propagationc}\\
\hat{p}_L^{\mathrm{out}}&=&\hat{p}_L^{\mathrm{in}}.\label{propagationd}
\end{eqnarray}
\endnumparts
The above equations can be straightforwardly generalized to the case in which a
single light beam $(\hat{x}_L,\hat{p}_L)$ propagates through many samples
shining at the $n$th sample at a certain angle $\alpha_n$.

Due to the strong polarization constraint the initial states of the atomic
ensembles as well as the one of light can be treated as Gaussian modes.
On the other hand, the interaction is a bilinear coupling between the Stokes
operator and the collective atomic spin operator, thus, it preserves the
Gaussian character of the initial modes and can be interpreted as a Gaussian
interaction between two bosonic modes. These facts enable us to tackle the
quantum interface within a CV formalism.

\section{ The matter-light interface in the CV formalism}

We start by reviewing the most basic concepts needed to describe Gaussian
continuous-variable systems. For further reading, the reader is referred to
\cite{Giedke2002_gaussops,Braunstein2005_rev,Adesso2007_rev} and references
therein. For a general quantum system of $N$ pairs of canonical degrees of
freedom (``position'' and ``momentum''), the commutation relations fulfilled by
the canonical coordinates $\hat{R}=(\hat{x}_1, \hat{p}_1, \ldots, \hat{x}_{N},
\hat{p}_{N})$ can be represented in a matrix form by the symplectic matrix
$\mathcal{J}_N: [\hat{R}_i,\hat{R}_j]=\im \hbar (\mathcal{J}_N)_{ij}$,
$i,j=1,\ldots, 2N$, where
\begin{equation}\label{sympl_J}
\mathcal{J}_{N}=\bigoplus_{\mu=1}^{N} \mathcal{J},\quad \mathcal{J}=\left(
                                                                      \begin{array}{cc}
                                                                        0 & 1 \\
                                                                        -1 & 0 \\
                                                                      \end{array}
                                                                    \right).
\end{equation}
Gaussian states are, by definition, fully described by the first and second
moments of the canonical coordinates. Hence, rather than describing them by
their infinite-dimensional density matrix $\varrho$, one can use the Wigner
function representation
\begin{equation}\label{wigner_gauss}
W(\zeta)=\frac{1}{\pi^N \sqrt{\det \gamma}} \exp \left[-(\zeta-d)^T \gamma^{-1} (\zeta-d)\right],
\end{equation}
which is a function of the first moments through the displacement
vector $d$, and of the second moments through the covariance
matrix $\gamma$, defined as:
\begin{equation}\label{disp_cov_def}
d_i=\Tr (\varrho \hat{R}_i),\qquad \gamma_{ij}=\Tr (\varrho \{\hat{R}_i-d_i,\hat{R}_j-d_j\}).
\end{equation}
The variable $\zeta=(x_1,p_1,\ldots,x_N,p_N)$ is a real phase space vector with
probability distribution given by the Wigner function. The covariance matrix
corresponding to a quantum state must fulfill the positivity condition
\begin{equation}\label{positivity}
\gamma+\im \mathcal{J}_N\geq 0.
\end{equation}
In the particular case of a physical system consisting of several atomic
ensembles and single light beam the most general covariance matrix takes the
form
\begin{equation}\label{cm_ABC}
\gamma=\left(
\begin{array}{cc}
  \gamma^{A} & C \\
  C^T & \gamma^{L}
\end{array}\right),
\end{equation}
where the submatrix $\gamma^L$ corresponding to light mode, $\gamma^A$ the
atomic ensembles, and $C$ accounts for the matter light correlations.

If a Gaussian state undergoes a unitary evolution preserving its Gaussian
character, which is the case in the physical systems under consideration, then
the respective transformation at the level of the covariance matrix is
represented by a symplectic matrix $S$ acting as
\begin{equation}\label{symplectic2}
\gamma_{\mathrm{out}}=S^{T}\gamma_{\mathrm{in}} S.
\end{equation}

We illustrate how to reconstruct the evolution of the covariance matrix from
the propagation equations (\ref{propagationa})--(\ref{propagationd}). One notes
that the variables describing the system after interaction are expressed as a
linear combination of the initial ones. Let us denote by $K$ the following
linear transformation $K: (\hat x_{A,n}^{\rm{out}},\hat p_{A,n}^{\rm{out}},\hat
x_{L}^{\rm{out}},\hat p_{L}^{\rm{out}})^{T}=K (\hat x_{A,n}^{\rm{in}},\hat
p_{A,n}^{\rm{in}},\hat x_{L}^{\rm{in}},\hat p_{L}^{\rm{in}})^{T}$, which can be
straightforwardly obtained from equations
(\ref{propagationa})--(\ref{propagationd}) and for a single atomic mode reads:

\begin{equation}
\left(
  \begin{array}{c}
    \hat{x}_A^{\mathrm{out}} \\
    \hat{p}_A^{\mathrm{out}} \\
    \hat{x}_L^{\mathrm{out}} \\
    \hat{p}_L^{\mathrm{out}}
  \end{array}
\right)= \left(
  \begin{array}{cccc}
    1 & 0& 0 & -\kappa \cos \alpha \\
    0 & 1 & 0 & \kappa \sin \alpha \\
    -\kappa \sin \alpha & -\kappa \cos \alpha & 1 & 0 \\
    0 & 0 & 0 & 1
  \end{array}
\right)\left(
  \begin{array}{c}
    \hat{x}_A^{\mathrm{in}} \\
    \hat{p}_A^{\mathrm{in}} \\
    \hat{x}_L^{\mathrm{in}} \\
    \hat{p}_L^{\mathrm{in}}
  \end{array}
\right).
\end{equation}
Since the interaction Hamiltonian is bilinear, the matrix $K$ can be directly
applied to a phase space vector $\zeta$ and correspondingly to the covariance
matrix, however the sign of the coupling constant $\kappa$ should be changed.
This is because the phase space variables undergo the Schr\"odinger evolution,
whereas the quadratures transform according to the Heisenberg picture.
Therefore, we define $\tilde{K}=K\big|_{\kappa\to(-\kappa)}$, which we apply to
the phase space vector and covariance matrix as
\begin{eqnarray}\label{symplectic}
\zeta_{\mathrm{out}}^{T}\gamma_{\mathrm{in}}^{-1} \zeta_{\mathrm{out}}&=&\zeta_{\mathrm{in}}^T \tilde{K}^T \gamma_{\mathrm{in}}^{-1} \tilde{K} \zeta_{\mathrm{in}}\nonumber\\
&=&\zeta_{\mathrm{in}}^T (\tilde{K}^{-1} \gamma_{\mathrm{in}} (\tilde{K}^{T})^{-1})^{-1} \zeta_{\mathrm{in}}\nonumber\\
&=&\zeta_{\mathrm{in}}^{T} \gamma_{\mathrm{out}}^{-1}\zeta_{\mathrm{in}},
\end{eqnarray}
leading to $S=(\tilde{K}^{T})^{-1}$. The above formalism has been explicitly
developed for a single sample and a single beam, but it easily generalizes to
an arbitrary number of atomic samples and beams, as well as to various
geometrical settings.

Finally, the last ingredient essential to describe the matter--light interface
at the level of the covariance matrix is the homodyne detection of light.
Assuming a zero initial displacement and covariance matrix of the form
(\ref{cm_ABC}), the measurement of the quadrature $\hat{x}_L$ with outcome
$\tilde{x}_L$ leaves the atomic system in a state described by a covariance
matrix \cite{Eisert2002_Gaussimpossible}
\begin{equation}\label{cm_measurement1}
\gamma^{A'}=\gamma^{A}-C (X\gamma^{L}X)^{-1} C^T,
\end{equation}
and displacement
\begin{equation}\label{cm_measurement2}
d_{A}=C (X\gamma^{L}X)^{-1} (\tilde{x}_L,0),
\end{equation}
where the inverse is understood as an inverse on the support whenever the
matrix is not of full rank and $X$ is a diagonal matrix with the same dimension
as $\gamma^{L}$ with diagonal entries $(1,0,1,0,\ldots,1,0)$.

Inherent to the matter--light interface is the analysis of the remaining atomic
samples once the light beam has been measured , in order to check the
correlations induced between them. With the CV formalism we have access to the
complete information provided by the covariance matrix and displacement vector
after interaction. This makes verification of entanglement amenable to
covariance matrix criteria (see also \cite{Madsen2004_Gauss,
Sherson2005_Gauss1}).

A structural separability test, which can be only applied when a full
covariance matrix is available, is the positive partial transposition (PPT)
test \cite{Horodecki1996_crit, Peres1996_crit}. For continuous-variable
systems, it corresponds to partial time reversal of the covariance matrix
\cite{Simon2000_pptCV}, i.e. the change of the sign of the momentum for chosen
modes. If the partially time reversed covariance matrix does not fulfill the
positivity condition (\ref{positivity}), the corresponding state is entangled.
This test, however, checks only for the bipartite entanglement. For Gaussian
states this criterion is necessary and sufficient for entanglement of $1\times
N$ modes.

Experimentally more convenient separability test is based on variances of
collective observables. It was provided for two-mode states in
\cite{Duan2000_sepCV} and generalized to many-mode states in
\cite{vanLoock2003_sepCVmult}. It states that if an $N$ mode state is
separable, then the sum of the variances of the following operators:
\begin{eqnarray}
\hat{u}=h_1 \hat{x}_{1}+\ldots+h_{N} \hat{x}_{N}\nonumber\\
\hat{v}=g_1 \hat{p}_{1}+\ldots+g_{N} \hat{p}_{N}
\end{eqnarray}
is bounded from below by a function of the coefficients
$h_1,\ldots,h_{N},g_1,\ldots,g_{N}$. Mathematically, the inequality is
expressed as
\begin{equation}\label{var_ineq}
\var{\hat{u}}+\var{\hat{v}}\geq f(h_1,\ldots,h_{N},g_1,\ldots,g_{N})\hbar,
\end{equation}
where
\begin{equation}
f(h_1,\ldots,h_{N},g_1,\ldots,g_{N})=\left|h_l g_l+\sum_{r\in
I}h_r g_r\right|+\left|h_m g_m+\sum_{s\in I'}h_s g_s\right|.
\end{equation}
In the above formula the two modes, $l$ and $m$, are distinguished and the
remaining ones are grouped into two disjoined sets $I$ and $I'$. The criterion
(\ref{var_ineq}) holds for all bipartite splittings of a state defined by the
sets of indices $\{l\}\cup I$ and $\{m\}\cup I'$. For two mode states, the
criterion becomes a necessary and sufficient entanglement test, however only
after the state is transformed into its standard form by local operations
\cite{Duan2000_sepCV}. This local transformations, however, are determined by
the form of the covariance matrix. In this sense, the knowledge of the full
covariance matrix is essential in order to determine whether the state is
entangled. Since in experiment we usually do not have access to this
information, we cannot assume that the criterion decides unambiguously about
separability.

\section{Symplectic and Covariance matrices for bipartite and multipartite entanglement}\label{examples}
Let us illustrate the versatility of the formalism in two distinct examples,
the first one deals with entanglement of two atomic samples generated and
manipulated by interaction and measurement of a light pulse. In the second one,
with larger number of atomic samples, we want to show how a cluster state can
be generated using the atom--light interface.

\subsection{Bipartite entanglement for mixed states: Generation and verification}

\begin{figure}[!t]
\center
  \psfrag{y}{$\!\! y$}\psfrag{z}{$z$}\psfrag{a}{-$\!\vec{J}_x$}\psfrag{b}{$\!\vec{J}_x$}\psfrag{c}{$\!\vec{B}$}\psfrag{d}{-\;$\!\!\!\vec{B}$}\includegraphics[width=0.35\textwidth]{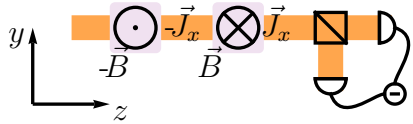}\\
  \caption{Schematically presented experimental setup used in \cite{Julsgaard2001_Nat} to
  generate and verify the presence of entanglement between two macroscopic atomic samples.
  The external magnetic field makes feasible the measurement of the two transverse components
  of the spin with a single light passage.}\label{bipPolzik}
\end{figure}

In the seminal work of Polzik and coworkers \cite{Julsgaard2001_Nat}, the
entanglement between two spatially separated atomic samples was generated in
the experimental setup schematically shown in figure \ref{bipPolzik}. In such
setup, both light and atomic samples were strongly polarized along the
$x$-direction while light propagated along the $z$-direction ($\alpha=0$). The
atomic ensembles were previously addressed with local magnetic fields oriented
in opposite directions, enabling to use a single light beam for generation of
$EPR$ entanglement and another one for verification, however, between new
time-integrated variables. It is straightforward to generalize the equations of
motion (\ref{propagationa})--(\ref{propagationd}) to two samples interacting
with light according to the Hamiltonian
\begin{equation}
\hat{H}_{\mathrm{int}}^{(2)}=-\frac{\kappa}{T} \hat{p}_L
(\hat{p}_{A,1}+\hat{p}_{A,2}).
\end{equation}
For example, in this case the canonical "momentum" variable for atoms,
$\hat{p}_{A,n}$ ($n=1,2$), is preserved [compare with equation
(\ref{propagationb})], while the canonical "position" of light after the
interaction carries information about the sum of atomic momenta
\begin{equation}\label{bipout}
 \hat{x}_L^{\rm{out}}=\hat{x}_L^{\rm{in}}-\kappa \left(\hat{p}_{A,1}+\hat{p}_{A,2}\right).
\end{equation}
Entanglement between the atomic samples is established as soon as the
$\hat{x}_L^{\rm{out}}$ component of light is measured. Moreover, it should be
emphasized that entanglement is generated independently of the outcome of the
measurement, nevertheless the measurement result indicates the displacement of
the state.

\begin{figure}[t]
\center
  \psfrag{y}{$\!\! y$}\psfrag{z}{$z$}\psfrag{a}{$\!\vec{J}_x$}\psfrag{b}{$\!\vec{J}_x$}
  a)\includegraphics[width=0.35\textwidth]{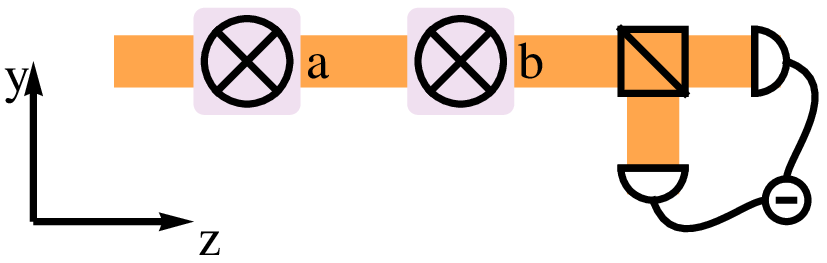}\quad b)\includegraphics[width=0.35\textwidth]{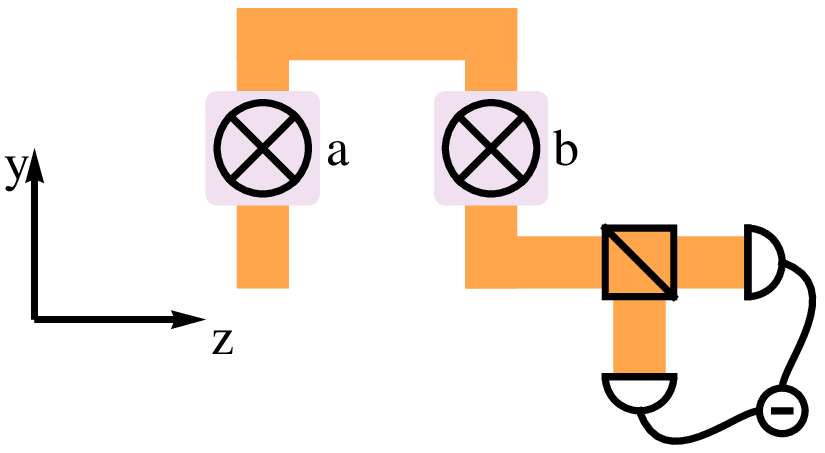}\\
  c)\includegraphics[width=0.35\textwidth]{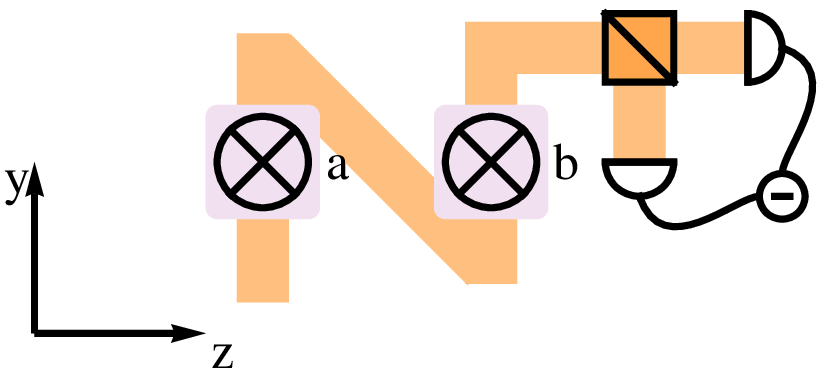}
  \caption{The sketch of the setups using the geometrical approach to generate
  and manipulate bipartite entanglement. The interaction between the light
  beam and atomic samples followed by the measurement introduces squeezing in
  a) $\hat{p}_{A,1}+\hat{p}_{A,2}$ b) $\hat{x}_{A,1}-\hat{x}_{A,2}$ and c) $\hat{x}_{A,1}+\hat{x}_{A,2}$.}\label{bipJulia}
\end{figure}

The setup described above needs some modifications if the individual magnetic
field addressing is impossible. The way to overcome this problem was shown in
\cite{Stasinska2009_meso}, and is summarized in figures \ref{bipJulia}a and
\ref{bipJulia}b.

Here we give a detailed description of this setup at the level of covariance
matrix. Moreover, we further assume that the initial state of the two samples
is not a minimum-fluctuation vacuum state as in \cite{Julsgaard_PhD}, but a
general thermal state. Under such assumptions, the initial state of the
composite system is given by the following covariance matrix for atoms and
light $\gamma_{\rm{in}}=n_1\id_{2}^{A}\oplus n_2\id_2^{A}\oplus\id_2^{L}$,
where the identity  $\id_{2}$ stands for a single mode and parameters $n_1,
n_2$ are related to temperature by $n_i=1/\tanh[\hbar \omega/(2 k_B T_i)]$
($i=1,2$). The symplectic matrix, $S_{\rm{int}}$, describing the interaction of
light passing through the samples at zero angle (figure \ref{bipJulia}a) is
given by

\begin{equation}
S_{\rm{int}}=\left(\begin{array}{cccc|cc}
 1 & 0 & 0 & 0 & 0 & 0\\
 0 & 1 & 0 & 0 & \kappa & 0\\
 0 & 0 & 1 & 0 & 0 & 0\\
 0 & 0 & 0 & 1 & \kappa & 0\\
 \hline
 0 & 0 & 0 & 0 & 1 & 0\\
 \kappa & 0 & \kappa & 0 & 0 & 1\\
 \end{array}\right),
\end{equation}
thus, the covariance matrix after the interaction takes the form expressed in
equation (\ref{symplectic2})
\begin{equation}
\gamma_{\rm{out}}=
\left(
\begin{array}{cccc|cc}
 n_1+\kappa ^2 & 0 & \kappa ^2 & 0 & 0 & \kappa  \\
 0 & n_1 & 0 & 0 & n_1 \kappa  & 0 \\
 \kappa ^2 & 0 & n_2+\kappa ^2 & 0 & 0 & \kappa  \\
 0 & 0 & 0 & n_2 & n_2 \kappa  & 0 \\
 \hline
 0 & n_1 \kappa  & 0 & n_2 \kappa  & 1+n_1 \kappa ^2+n_2 \kappa ^2 & 0 \\
 \kappa  & 0 & \kappa  & 0 & 0 & 1
\end{array}
\right).
\end{equation}
Both modes, representing the samples, are entangled with light, however their
reduced state is separable as one can check applying PPT criterion to the
covariance matrix of the upper-left block matrix. Entanglement between atomic
samples is not produced until one measures a quadrature of light. To
demonstrate this, we apply a homodyne measurement of the light mode [see
(\ref{cm_measurement1}) and (\ref{cm_measurement2})]. Assuming the measurement
outcome $\tilde{x}_{L,1}$ obtaining the covariance matrix describing the final
state of the samples

\begin{eqnarray}\label{cm_epr}
\gamma_{\rm{fin}}&=&\left(
\begin{array}{cccc}
 n_1+\kappa ^2 & 0 & \kappa ^2 & 0 \\
 0 & \frac{n_1 n_2 \kappa ^2+n_1}{\left(n_1+n_2\right) \kappa ^2+1} & 0 & -\frac{n_1 n_2 \kappa ^2}{\left(n_1+n_2\right) \kappa ^2+1} \\
 \kappa ^2 & 0 & n_2+\kappa ^2 & 0 \\
 0 & -\frac{n_1 n_2 \kappa ^2}{\left(n_1+n_2\right) \kappa ^2+1} & 0 & \frac{n_1 n_2 \kappa ^2+n_2}{\left(n_1+n_2\right) \kappa ^2+1}
\end{array}
\right),
\end{eqnarray}
and the displacement of the final state is
\begin{equation}\label{displ_epr}
d_{\mathrm{fin}}=\left(0,\frac{\tilde{x}_{L,1} \kappa }{2 \kappa
^2+1},0,\frac{\tilde{x}_{L,1} \kappa }{2 \kappa ^2+1}\right).
\end{equation}
Notice that the covariance matrix is independent of the measurement outcome,
but the latter is present in the displacement vector.

In order to verify that the final state of atomic samples is entangled we use
the separability criterion based on the variances of the two commuting
variables \cite{Duan2000_sepCV}
\begin{equation}\label{crit}
\varsb{(|\lambda|\hat{p}_{A,1}+\frac{1}{\lambda}\hat{p}_{A,2})}+
\varsb{(|\lambda|\hat{x}_{A,1}-\frac{1}{\lambda}\hat{x}_{A,2})}\geq 2\hbar.
\end{equation}
We emphasize, however, that due to the fact that in the analyzed setup the
measurement possibilities are limited, the only experimentally feasible case is
$|\lambda|=1$. The way to measure such combination of variances was described
in detail in \cite{Stasinska2009_meso}, therefore we will not recall it here.
Also the final state is not, in general, in its standard form, therefore, the
measurement gives only a sufficient condition for separability.

\begin{figure}[!t]
  \psfrag{n}{$n_1$}\psfrag{m}{$n_2$}\psfrag{k}{$\!\!\!\!\!\!\!\! \kappa$}\psfrag{f}{$1$}\psfrag{0}{$0$}\psfrag{1}{\ \\[2ex] $1$}\psfrag{2}{$2$}\psfrag{3}{$3$}\psfrag{4}{$4$}\psfrag{a}{$\!\!\!0.5$}\psfrag{b}{$\!\!\! 1.0$}\psfrag{c}{$\!\!\! 1.5$}\psfrag{d}{$\!\!\! 2.0$}
  a)\includegraphics[width=0.45\textwidth]{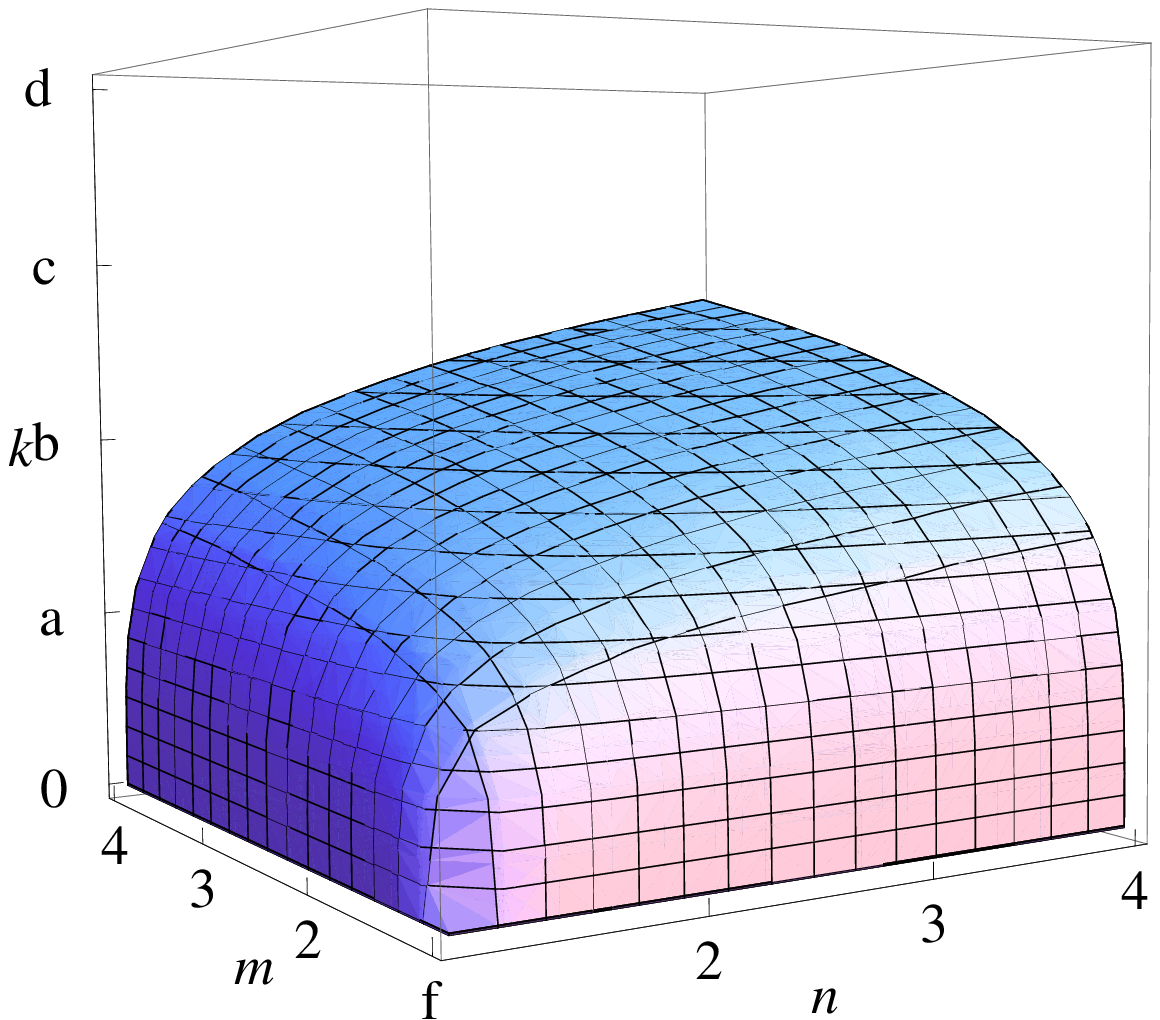}\quad b) \includegraphics[width=0.45\textwidth]{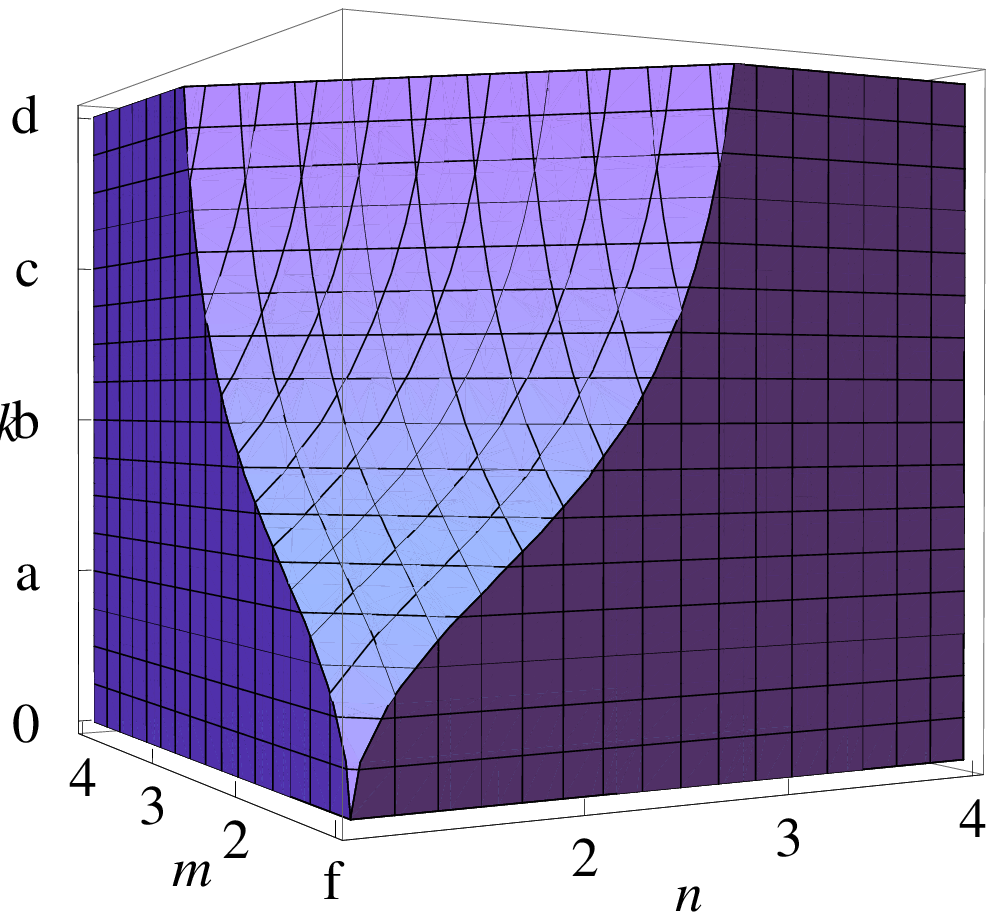}\\
  c)\includegraphics[width=0.45\textwidth]{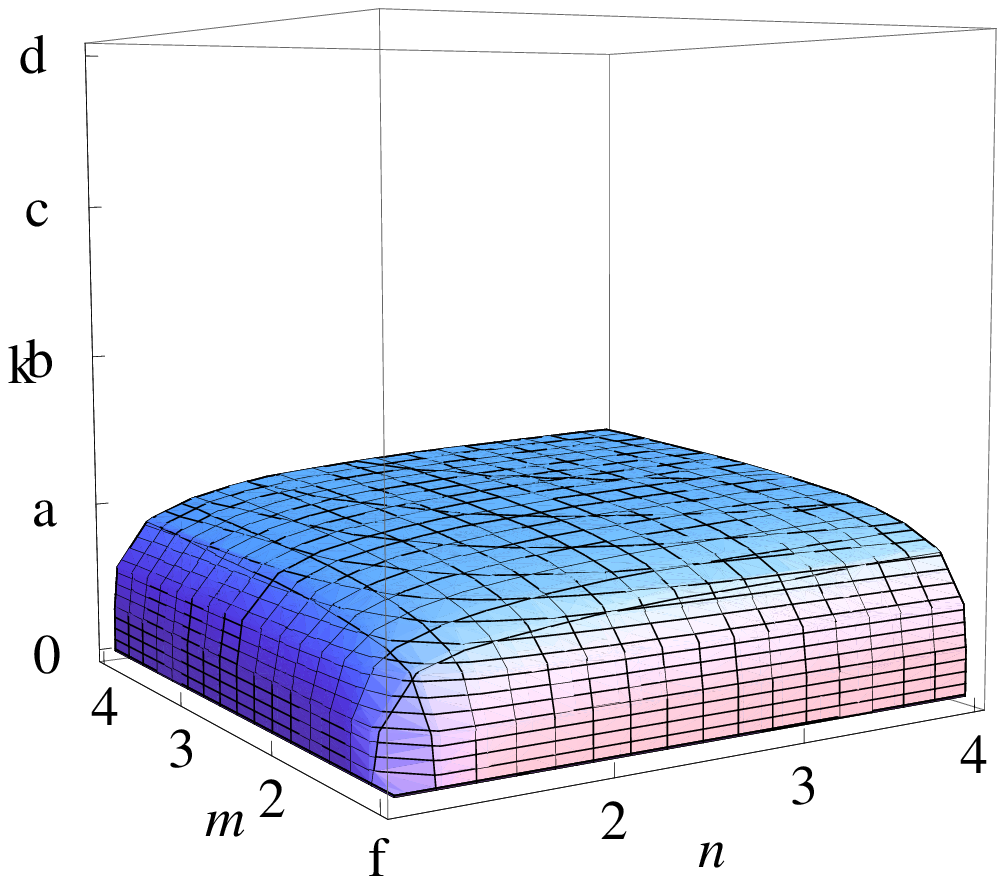}\quad d) \includegraphics[width=0.45\textwidth]{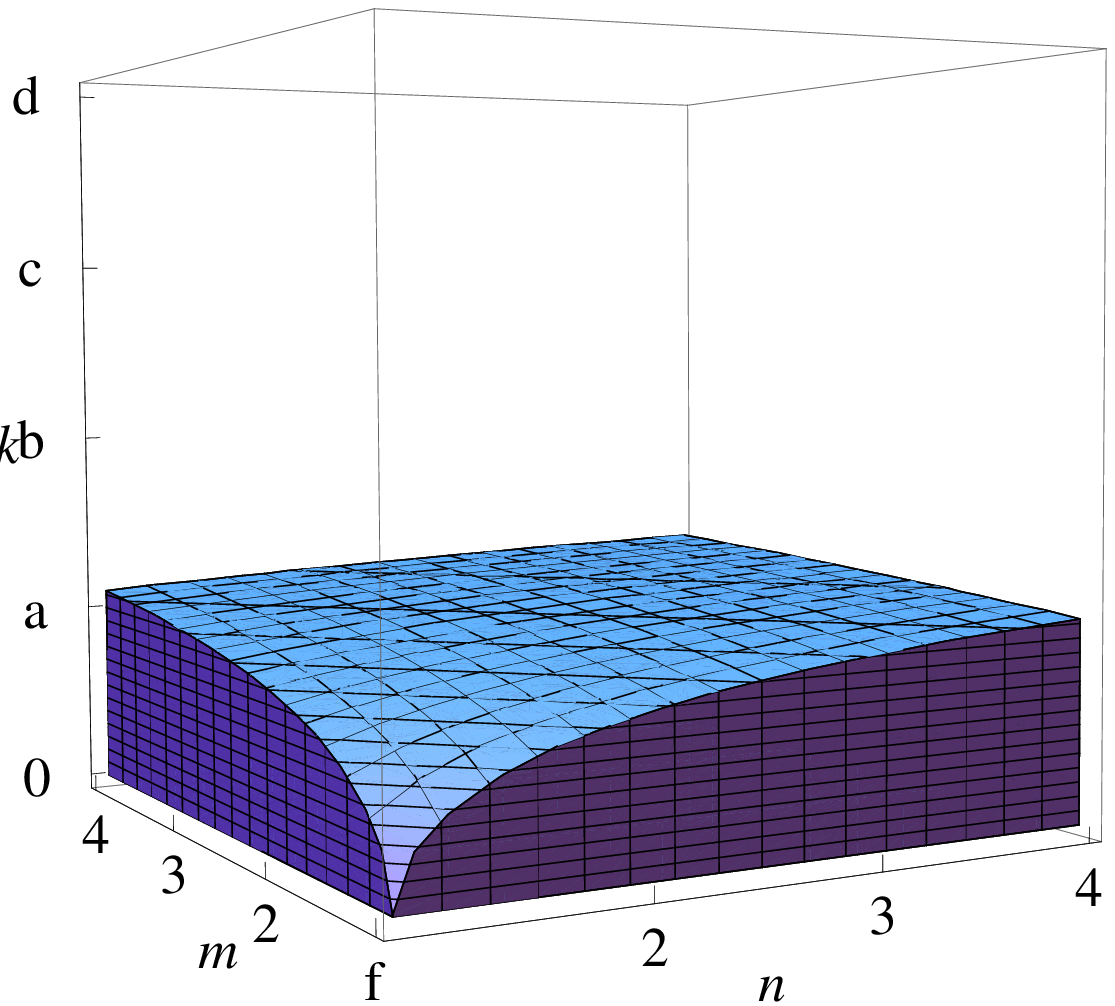}
  \caption{Comparison of the sets of parameters $\kappa, n_1, n_2$ for which the states are entangled [figures a) and
  c)], with the ranges of parameters for which they are detected by the spin variance inequality (\ref{crit}) with $\lambda=1$
  [figures b) and d)]. In figures a) and b) we consider the state produced in the setup from figure
  \ref{bipJulia}a, whereas in figures c) and d) the one produced in two steps schematically depicted in figures \ref{bipJulia}a and
  \ref{bipJulia}b.}\label{ppt_thermal}
\end{figure}

In the following analysis we restrict to a single inequality involving the
collective observables which are squeezed during entanglement generation, i.e.
the one for $\lambda=1$.
From the final covariance matrix (\ref{cm_epr}) one can directly
compute the variances of the collective atomic spin:
\begin{eqnarray}
\frac{1}{\hbar}\varsb{(\hat{p}_{A,1}+\hat{p}_{A,2})}&=&\frac{1}{2} (\gamma_{\mathrm{fin},22}+\gamma_{\mathrm{fin},44}+2 \gamma_{\mathrm{fin},24})\nonumber\\
&=&\frac{n_1+n_2}{2 \left(n_1+n_2\right) \kappa ^2+2},\nonumber\\
\frac{1}{\hbar}\varsb{(\hat{x}_{A,1}-\hat{x}_{A,2})}&=&\frac{1}{2} (\gamma_{\mathrm{fin},11}+\gamma_{\mathrm{fin},33}-2 \gamma_{\mathrm{fin},13})\nonumber\\
&=&\frac{1}{2} \left(n_1+n_2\right).
\end{eqnarray}
Substituting the obtained variances into the separability criterion we obtain
\begin{equation}
\varsb{(\hat{p}_{A,1}+\hat{p}_{A,2})}+\varsb{(\hat{x}_{A,1}-\hat{x}_{A,2})}=\frac{(n_1+n_2)}{2}
\frac{\left(n_1+n_2\right) \kappa ^2+2}{\left(n_1+n_2\right) \kappa ^2+1},
\end{equation}
which violates the bound $2\hbar$ for specific values of $n_1,n_2$, and
$\kappa$, i.e. for those that fulfill
\begin{equation}\label{var_thermal}
\kappa>\frac{\sqrt{2} \sqrt{n_1+n_2-2}}{\sqrt{\left(-n_1-n_2+4\right)
\left(n_1+n_2\right)}}.
\end{equation}
The criterion (\ref{crit}) for $\lambda=1$ does not detect all entanglement. We
compare the set that is not detected by the inequality to the set of separable
states found using the PPT criterion. The respective ranges of parameters
$\kappa,n_1,n_2$ are depicted in figures \ref{ppt_thermal}a and
\ref{ppt_thermal}b.

The comparison shows that to generate entanglement between thermal states, much
stronger coupling is necessary. Also, the simple criterion based on the
variances cannot detect entanglement for $n_1+n_2>4$, even though it is clearly
present in the system.

The procedure summarized in figure \ref{bipJulia}b introduces more entanglement
in the system squeezing the collective variable $\hat{x}_{A,1}-\hat{x}_{A,2}$.
The computation is analogous, therefore will not be repeated. The resulting
state is characterized by the following variances:
\begin{eqnarray}
\frac{1}{\hbar}\varsb{(\hat{p}_{A,1}+\hat{p}_{A,2})}=\frac{1}{\hbar}\varsb{(\hat{x}_{A,1}-\hat{x}_{A,2})}=\frac{n_1+n_2}{2
\left(n_1+n_2\right) \kappa ^2+2}.
\end{eqnarray}
The condition for the inequality (\ref{crit}) for $\lambda=1$ to be violated is
\begin{equation}
\kappa>\sqrt{\frac{n_1+n_2-2}{2 n_1+2 n_2}}.
\end{equation}
The set of parameters $\kappa, n1, n2$ for which the final state is detected by
the inequality is depicted in figure \ref{ppt_thermal}c. For comparison the set
of separable (equivalently PPT) is shown in figure \ref{ppt_thermal}d. Again
the inequality does not detect all entangled states, however is more efficient
than in the first case.


Interestingly enough, our geometrical approach, making the light impinging on
each atomic sample at a given angle $\alpha_i$, also opens the possibility of
deleting all the entanglement created by the first light beam, by interaction
with a second light beam of an appropriate intensity. Notice that the
entanglement procedure is intrinsically irreversible because it involves a
projective measurement, so coming ``deterministically'' back to the initial
state is not obvious. In \cite{Filip2002_CVerasing,Filip2003_CVerasing}, a
quantum erasing scheme in continuous-variable systems was proposed. The
measurement of the meter coordinate entangled with the quantum system causes a
back-action. The authors shown that it is possible to erase the action of the
measurement and restore the original state of the system. Here we are
interested in deleting the measurement induced entanglement between two atomic
samples, exploiting the squeezing and antisqueezing effects produced by the
laser beams. Again, a CV formalism greatly simplifies the analysis and
indicates in which way the atomic quadratures, squeezed and anti-squeezed by
the first light beam, can be restored with the help of a second light pulse.

We begin with the state produced in the first part of the section, represented
by the covariance matrix (\ref{cm_epr}). We will perform this part of the
analysis only for the initially vacuum state, i.e. $n_1=n_2=1$. Evolution due
to the interaction with the light beam impinging on each atomic sample at
$\alpha_1=\alpha_2=\pi/2$, schematically represented in figure \ref{bipJulia}c
is reproduced by the following symplectic matrix acting on $\gamma_{\rm
fin}\oplus \mathbbm{1}_2$:
\begin{equation}
S_{\rm int}^{\rm eraser}=\left(
\begin{array}{cccc|cc}
 1 & 0 & 0 & 0 & \eta  & 0 \\
 0 & 1 & 0 & 0 & 0 & 0 \\
 0 & 0 & 1 & 0 & \eta  & 0 \\
 0 & 0 & 0 & 1 & 0 & 0 \\
 \hline
 0 & 0 & 0 & 0 & 1 & 0 \\
 0 & -\eta  & 0 & -\eta  & 0 & 1
\end{array}
\right).
\end{equation}
In general, the second light beam may have different properties, hence we use a
different coupling constant $\eta$. After the interaction and measurement of
light the state of atomic ensembles is characterized by the following
covariance matrix:
\begin{equation}\label{cm_erased}
\gamma_{\rm 2nd}=\left(
\begin{array}{cccc}
 \frac{\eta ^2 \left(2 \kappa ^2+1\right)+\kappa
   ^2+1}{2 \eta ^2 \left(2 \kappa ^2+1\right)+1} & 0 &
   \frac{\kappa ^2-\eta ^2 \left(2 \kappa
   ^2+1\right)}{2 \eta ^2 \left(2 \kappa ^2+1\right)+1}
   & 0 \\
 0 & \frac{\eta ^2 \left(2 \kappa ^2+1\right)+\kappa
   ^2+1}{2 \kappa ^2+1} & 0 & -\frac{\kappa ^2-\eta ^2
   \left(2 \kappa ^2+1\right)}{2 \kappa ^2+1} \\
 \frac{\kappa ^2-\eta ^2 \left(2 \kappa ^2+1\right)}{2
   \eta ^2 \left(2 \kappa ^2+1\right)+1} & 0 &
   \frac{\eta ^2 \left(2 \kappa ^2+1\right)+\kappa
   ^2+1}{2 \eta ^2 \left(2 \kappa ^2+1\right)+1} & 0 \\
 0 & -\frac{\kappa ^2-\eta ^2 \left(2 \kappa
   ^2+1\right)}{2 \kappa ^2+1} & 0 & \frac{\eta ^2
   \left(2 \kappa ^2+1\right)+\kappa ^2+1}{2 \kappa
   ^2+1}
\end{array}
\right).
\end{equation}
For the specifically adjusted values of $\kappa$ and $\eta$, i.e.
$\eta^2=\kappa^2/(1+2\kappa^2)$, the atomic ensembles come back to the initial
coherent state, at least at the level of variances. Let us have a look at the
displacement vector. Assuming that the outcome of the measurement leading to
the generation of entangled state (\ref{cm_epr}) is $\tilde{x}_{L,1}$ (see
equation (\ref{displ_epr})), and the outcome of the measurement erasing the
correlations is $\tilde{x}_{L,2}$, the total displacement is given by
\begin{equation}
d_{A}=\left(\sqrt{\frac{\kappa ^2}{2 \kappa
   ^2+1}}\tilde{x}_{L,2},\frac{\kappa }{2 \kappa
   ^2+1}\tilde{x}_{L,1},\sqrt{\frac{\kappa ^2}{2 \kappa
   ^2+1}}\tilde{x}_{L,2},\frac{\kappa }{2 \kappa
   ^2+1}\tilde{x}_{L,1}\right).
\end{equation}
Hence the final state is equivalent to the initial one only up to the
displacement vector. 

\subsection{Multipartite entanglement: Generation of cluster states using atom-light interfaces. }

In \cite{Briegel2001_cluster} a class of $N$-qubit quantum states generated in
two dimensional arrays of qubits using an Ising-type interaction was presented;
these are the so-called cluster states. Using their scalability properties,
Briegel {\it et al.} proposed a scheme for a one-way quantum computation
\cite{Raussendorf2001_oneway,Raussendorf2003_onewayPRA}. There, two-dimensional
cluster states are the entire resource, while computation consists of a
sequence of local projective measurements. Furthermore, any two-dimensional
cluster state provides a universal quantum computer since it has been proved
that any unitary quantum logic network can be efficiently simulated within this
scheme. In this sense cluster states, can be regarded as a resource able to
generate any type of multi-qubit entanglement by means of two body
interactions.

We present here, how the continuous-variable cluster-like states
\cite{vanLoock2007_CVcluster,vanLoock2008_CVcluster4} can be generated within
the analyzed atom--light interface. We associate the modes of the $N_s$-mode
system with the vertices of a graph $G$. The edges between the vertices define
the notion of nearest neighbourhood. By $N_a$ we denote the set of nearest
neighbours of vertex $a$. A cluster is a connected graph. For angular momentum
variables, cluster states are defined only asymptotically as those with
infinite squeezing in the variables
\begin{equation}
\hat{p}_{a}-\sum_{b\in N_a} \hat{x}_{b},
\end{equation}
for all $a\in G$. We talk about the clusterlike states when
squeezing is finite.

\begin{figure}
\center
  \psfrag{y}{$\!\!
  y'$}\psfrag{z}{$\!z'$}\psfrag{a}{$\!\vec{J}_x$}\psfrag{b}{$\!\vec{J}_x$}\psfrag{c}{$\!\vec{J}_x$}\psfrag{d}{$\!\vec{J}_x$}\psfrag{1}{$1$}\psfrag{2}{$2$}\psfrag{3}{$3$}\psfrag{4}{$4$}
  a)\includegraphics[width=0.4\textwidth]{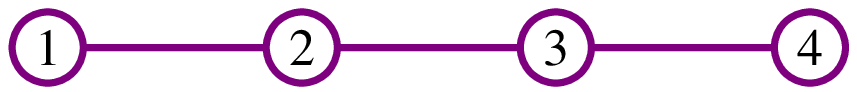}\\[2ex]
  b)\includegraphics[width=0.49\textwidth]{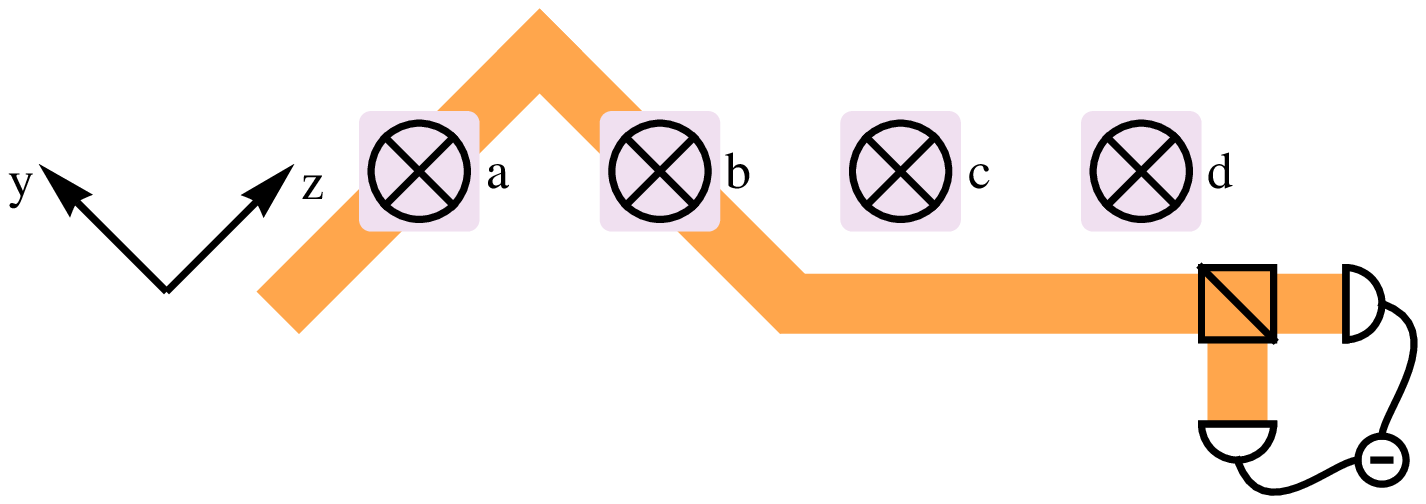}c)\includegraphics[width=0.49\textwidth]{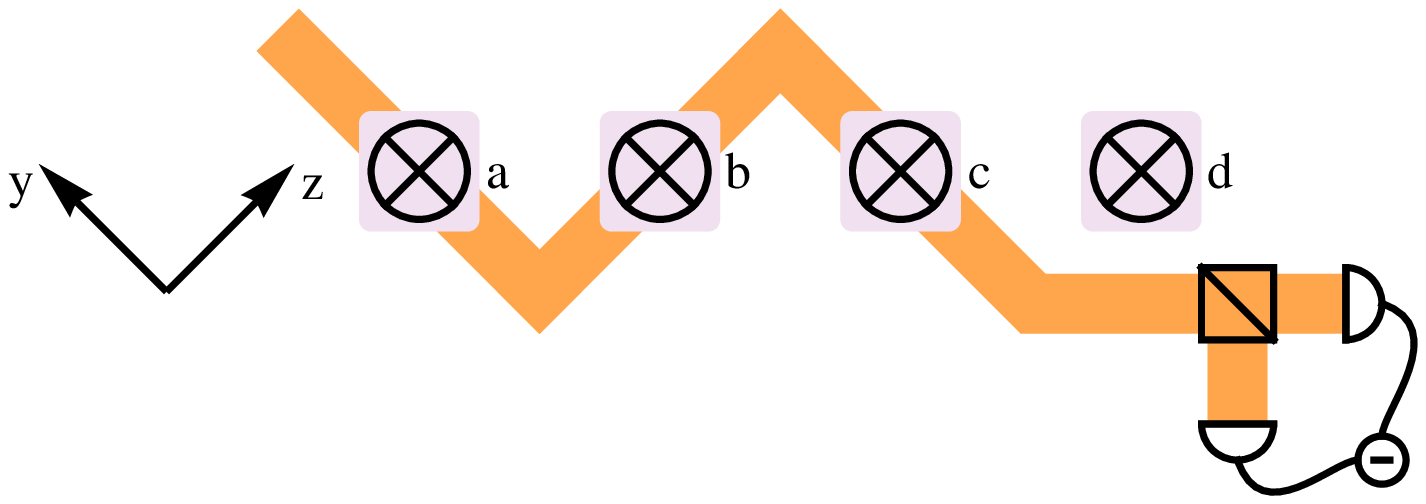}\\
  d)\includegraphics[width=0.49\textwidth]{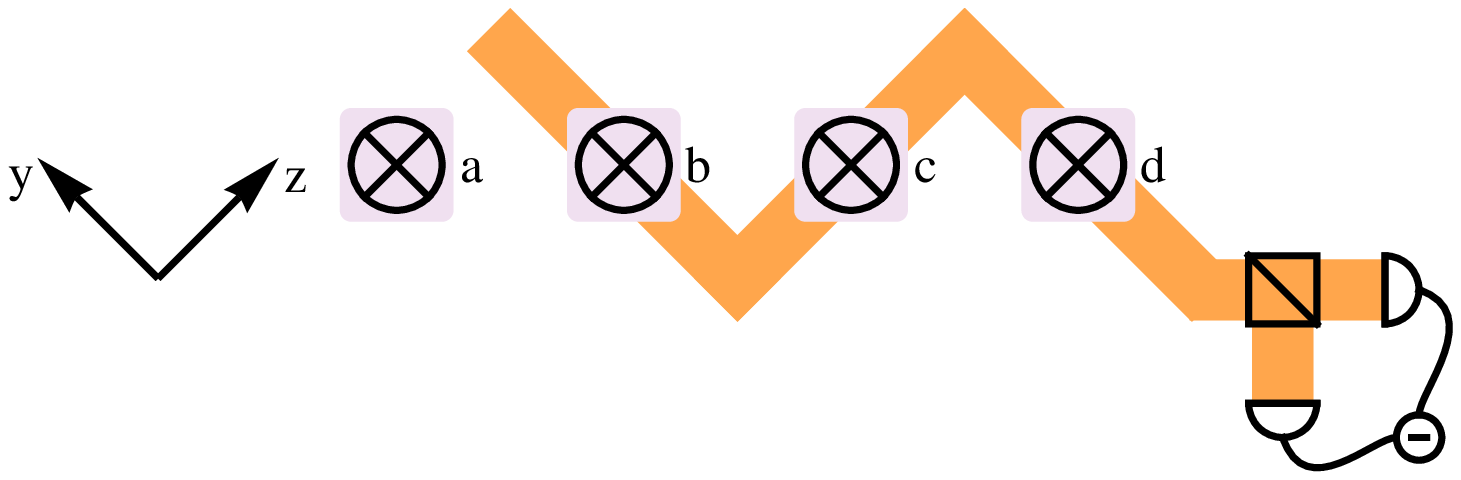}e)\includegraphics[width=0.49\textwidth]{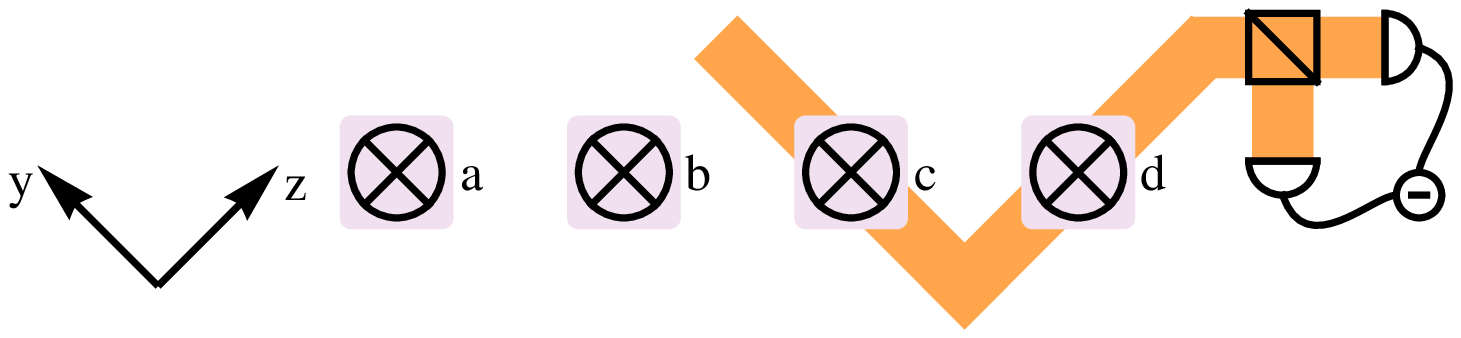}\\
  \caption{Schematically depicted setups leading to generation of the four-partite linear cluster state.
  Figure a) shows the graph representing the state structure. In figures b)-e) the sequence of beams
  introducing the CV cluster-like corelations between atomic samples are depicted: b) $\hat{p}_{A,1}'-\hat{x}_{A,2}'$,
  c) $\hat{p}_{A,2}'-\hat{x}_{A,1}'-\hat{x}_{A,3}'$, d) $\hat{p}_{A,3}'-\hat{x}_{A,2}'-\hat{x}_{A,4}'$
  and e) $\hat{p}_{A,4}'-\hat{x}_{A,3}'$.}\label{fig_cluster}
\end{figure}

In \cite{Stasinska2009_meso} we have shown how to generate a simple
cluster-like state between four samples (figure \ref{fig_cluster}a), however in
a rotated frame. For each mode we introduce new variables:
\begin{equation}
\hat{x}_{A,n}'=\frac{1}{\sqrt{2}} (\hat{x}_{A,n}-\hat{p}_{A,n}), \quad
\hat{p}_{A,n}'=\frac{1}{\sqrt{2}} (\hat{x}_{A,n}+\hat{p}_{A,n}).
\end{equation}
The setup is summarized in figure \ref{fig_cluster}.

The state is generated in four, commuting steps. We will denote an initial
state of light and atoms at each step by $\gamma_{\mathrm{in}}^{(j)}$, a state
after interaction by $\gamma_{\mathrm{out}}^{(j)}$, and the state after the
measurement of light by $\gamma_{\mathrm{fin}}^{(j)}$ $j=1,\ldots,4$. Note that
$\gamma_{\mathrm{in}}^{(j+1)}=\gamma_{\mathrm{fin}}^{(j)}\oplus \mathbbm{1}_2$.

We assume that initially both the atomic samples and light are in a vacuum
state $\gamma_{\mathrm{in}}^{(1)}=\mathbbm{1}_{8+2}$. The symplectic matrix (in
a rotated frame) reproducing the effect of interaction depicted in figure
\ref{fig_cluster}b is
\begin{equation}
S_I^{(1)}=\left(
  \begin{array}{c|c|c}
    \begin{array}{ccc}
    1 & & 0 \\
    0 & & 1
    \end{array} &
    \begin{array}{ccc}
    & \boldsymbol{0} &
    \end{array} &
    \begin{array}{cc}
    0 & 0 \\
    \kappa & 0
    \end{array} \\
    \hline
    \begin{array}{c}
     \\
     \\
    \boldsymbol{0} \\
     \\
     \\
    \end{array} &
    \begin{array}{ccccc}
     & & & & \\
     & & & & \\
     & & \mathbbm{1} & & \\
     & & & & \\
     & & & &
    \end{array} &
    \begin{array}{cc}
    -\kappa & 0 \\
    0 & 0 \\
    & \\
    \quad\, \boldsymbol{0} & \\
    &
    \end{array} \\
    \hline
    \begin{array}{cc}
    0 & 0   \\
    \kappa & 0
    \end{array} &
    \begin{array}{cc}
    0 & 0   \\
    0 & k
    \end{array}
    \begin{array}{ccc}
    & \boldsymbol{0} &
    \end{array} &
    \begin{array}{ccc}
    1 & & 0   \\
    0 & & 1
    \end{array}
  \end{array}
\right),
\end{equation}
whereas the interaction from figure \ref{fig_cluster}c is reproduced with
symplectic matrix
\begin{equation}
S_I^{(2)}=\left(
  \begin{array}{c|c|c}
    \begin{array}{ccc}
    1 & & 0 \\
    0 & & 1
    \end{array} &
    \begin{array}{ccc}
    & \boldsymbol{0} &
    \end{array} &
    \begin{array}{cc}
    -\kappa & 0 \\
    0 & 0
    \end{array} \\
    \hline
    \begin{array}{c}
     \\
     \\
    \boldsymbol{0} \\
     \\
     \\
    \end{array} &
    \begin{array}{ccccc}
     & & & & \\
     & & & & \\
     & & \mathbbm{1} & & \\
     & & & & \\
     & & & &
    \end{array} &
    \begin{array}{cc}
    0 & 0 \\
    \kappa & 0 \\
    -\kappa & 0 \\
    0 & 0 \\
    0 & 0\\
    0 & 0
    \end{array} \\
    \hline
    \begin{array}{cc}
    0 & 0   \\
    0 & \kappa
    \end{array} &
    \begin{array}{cccccc}
    0 & 0 & 0 & 0 & 0 & 0\\
    \kappa & 0 & 0 & \kappa & 0 & 0
    \end{array}&
    \begin{array}{ccc}
    1 & & 0   \\
    0 & & 1
    \end{array}
  \end{array}
\right).
\end{equation}
The other two symplectic matrices can be easily written down
basing on $S_I^{(1)}$ and $S_I^{(2)}$.

The final state after the sequence of four atom-light interactions
followed by the measurement of light is characterized by the
covariance matrix
\begin{equation}\label{cm_cluster}
\gamma_{\mathrm{fin}}^{(4)}=
\left(
\begin{array}{cccc}
A & C & E & F\\
C & B & D & E\\
E & D & B & C\\
F & E & C & A
\end{array}
\right).
\end{equation}
where $A,B,C,D,E$ and $F$ are the following $2\times 2$ matrices:
\begin{eqnarray}
\hspace{-2.5cm}\begin{array}{l} A=\left(\!
\begin{array}{cc}
 \frac{5 \kappa ^6+8 \kappa ^4+5 \kappa ^2+1}{5 \kappa ^4+5 \kappa ^2+1}\!\!\! & 0 \\
 0 \!\!\!& \frac{5 \kappa ^6+7 \kappa ^4+5 \kappa ^2+1}{5 \kappa ^4+5 \kappa ^2+1}
\end{array}
\!\right),\\
B=\left(\!
\begin{array}{cc}
 \frac{5 \kappa ^6+7 \kappa ^4+4 \kappa ^2+1}{5 \kappa ^4+5 \kappa ^2+1}\!\!\! & 0 \\
 0\!\!\! & \frac{10 \kappa ^6+13 \kappa ^4+6 \kappa ^2+1}{5 \kappa ^4+5 \kappa ^2+1}
\end{array}
\!\right),\\
C=\left(\!
\begin{array}{cc}
 0 \!\!\!& \frac{\kappa ^2 \left(5 \kappa ^4+7 \kappa ^2+2\right)}{5 \kappa ^4+5 \kappa ^2+1} \\
 \frac{\kappa ^2 \left(5 \kappa ^4+7 \kappa ^2+2\right)}{5 \kappa ^4+5 \kappa ^2+1}\!\!\! & 0
\end{array}
\!\right),
\end{array} \quad
\begin{array}{l}
D=\left(\!
\begin{array}{cc}
 0 \!\!\! & \frac{\kappa ^2 \left(5 \kappa ^4+6 \kappa ^2+2\right)}{5 \kappa ^4+5 \kappa ^2+1} \\
 \frac{\kappa ^2 \left(5 \kappa ^4+6 \kappa ^2+2\right)}{5 \kappa ^4+5 \kappa ^2+1} \!\!\!& 0
\end{array}
\!\right),\\
E=\left(\!
\begin{array}{cc}
 \frac{-\kappa ^4-\kappa ^2}{5 \kappa ^4+5 \kappa ^2+1}\!\!\! & 0 \\
 0\!\!\! & \frac{5 \kappa ^6+6 \kappa ^4+\kappa ^2}{5 \kappa ^4+5 \kappa ^2+1}
\end{array}
\!\right),\\
F=\left(\!
\begin{array}{cc}
 0\!\!\! & -\frac{\kappa ^4}{5 \kappa ^4+5 \kappa ^2+1} \\
 -\frac{\kappa ^4}{5 \kappa ^4+5 \kappa ^2+1}\!\!\! & 0
\end{array}
\!\right).
\end{array}
\end{eqnarray}
In order to check the separability of the produced state, we apply
the PPT criterion. The partial time reversal of covariance matrix
(\ref{cm_cluster}) with respect to all cuts is negative.
Therefore, the state is fully inseparable for all values of
$\kappa$.

\begin{figure}[!h]
\center
  \psfrag{k}{$\;\kappa$}\psfrag{0.5}{$\!\!0.5$}\psfrag{0.4}{$\!0.4$}\psfrag{0.2}{$\!0.2$}\psfrag{0.8}{$\!0.8$}\psfrag{2.5}{$\!\!2.5$}\psfrag{3.0}{$\!\!3.0$}\psfrag{2.0}{$\!\!2.0$}\psfrag{1.5}{$\!\!\!1.5$}\psfrag{1.0}{$\!\!\!1.0$}\psfrag{0.6}{$\!0.6$}\includegraphics[width=0.4\textwidth]{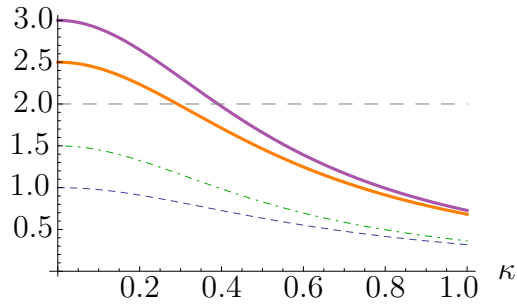}\\
  \caption{Dashed line - equation (\ref{plot1}), dashed-dotted line - equation (\ref{plot2}), purple - equation (\ref{plot3}), orange - equation (\ref{plot4}) and (\ref{plot5})}\label{fig_plots}
\end{figure}

Applying a less effective, but experimentally convenient separability test of
the form (\ref{var_ineq}) one detects full inseparability above some threshold
value of $\kappa$. The three inequalities which constitute a necessary
separability criterion for this particular cluster state contain variances of
the variables squeezed by the interaction and read
\cite{vanLoock2008_CVcluster4}:
\numparts
\begin{eqnarray}\label{ineq_cluster}
&&\Delta_1\equiv\varsb{(\hat{p}'_{1}-\hat{x}'_{2})}+\varsb{(\hat{p}'_{2}-\hat{x}'_{1}-\hat{x}'_{3})}\geq 2\hbar\label{plot4}\\
&&\Delta_2\equiv\varsb{(\hat{p}'_{3}-\hat{x}'_{2}-\hat{x}'_{4})}+\varsb{(\hat{p}'_{2}-\hat{x}'_{1}-\hat{x}'_{3})}\geq 2\hbar\label{plot3}\\
&&\Delta_3\equiv\varsb{(\hat{p}'_{2}-\hat{x}'_{1}-\hat{x}'_{3})}+\varsb{(\hat{p}'_{4}-\hat{x}'_{3})}\geq\label{plot5}
2\hbar.
\end{eqnarray}
\endnumparts
We determine the variances explicitly from the covariance matrix
(\ref{cm_cluster}):
\numparts
\begin{eqnarray}
\frac{1}{\hbar}\varsb{(\hat{p}'^{(1)}-\hat{x}'^{(2)})}&=&\frac{1}{2} A_{22}+\frac{1}{2} B_{11}-C_{21}\nonumber\\
&=&\frac{5 \kappa ^2+2}{10 \kappa ^4+10 \kappa ^2+2},\label{plot1}
\end{eqnarray}
\begin{eqnarray}
\frac{1}{\hbar}\varsb{(\hat{p}'^{(2)}-\hat{x}'^{(1)}-\hat{x}'^{(3)})}&=&\frac{1}{2}
B_{22}+\frac{1}{2} A_{11}+\frac{1}{2}
B_{11}-C_{12}-D_{21}+E_{11}\nonumber\\&=&\frac{5 \kappa ^2+3}{10 \kappa ^4+10
\kappa ^2+2},\label{plot2}
\end{eqnarray}
\begin{eqnarray}
\frac{1}{\hbar}\varsb{(\hat{p}'^{(3)}-\hat{x}'^{(2)}-\hat{x}'^{(4)})}&=&\frac{1}{2} B_{22}+\frac{1}{2} B_{11}+\frac{1}{2} A_{11}-D_{12}-C_{21}+E_{11}\nonumber\\
&=&\frac{5 \kappa ^2+3}{10 \kappa ^4+10 \kappa ^2+2},
\end{eqnarray}
\begin{eqnarray}
\frac{1}{\hbar}\varsb{(\hat{p}'^{(4)}-\hat{x}'^{(3)})}&=&\frac{1}{2}A_{22}+\frac{1}{2}B_{11}-C_{12}\nonumber\\
&=&\frac{5\kappa ^2+2}{10 \kappa ^4+10 \kappa ^2+2}.
\end{eqnarray}
\endnumparts
The dependence of the left-hand sides of inequalities (\ref{ineq_cluster}) on
the coupling constant is depicted in figure \ref{fig_plots}. The inequalities
(\ref{plot4}) and (\ref{plot5}) are violated for $\kappa\gtrsim 0.3 $ and the
inequality (\ref{plot3}) is violated for $\kappa\gtrsim 0.4$.
%

\section{Summary}\label{summary}
We have addressed the problem of manipulation of entanglement between atomic
ensembles using Faraday quantum atom--light interface with a
continuous-variable formalism. Both the atomic ensembles and light can be
prepared in initial Gaussian state and remain Gaussian after all the steps
leading to generation of entanglement. This allows us to describe the whole
process using the covariance matrix. After a general introduction of the
mentods we have applied it to derive bounds on the strength of the interaction
leading to entanglement between atomic ensembles prepared initially in mixed
states. Then we have shown how the CV formalism facilitates the analysis of
more complex systems. In particular we have addressed the problem of generation
of the CV cluster states.

\ack The authors acknowledge support from the Spanish MICINN Grant No.
FIS2008-01236, Generalitat de Catalunya Grant No. 2005SGR-00343 and Consolider
Ingenio 2010 QOIT. J.S. is supported by the Spanish Ministry of Education
through the program FPU. S.P. is supported by the Spanish Ministry of Science
and Innovation through the program Juan de la Cierva.

\section*{References}
\bibliography{methods_iop}

\end{document}